\documentclass[12pt]{article}
\usepackage{indentfirst} 
\usepackage{amsfonts,amssymb,amsmath}
\usepackage{bm}          
\usepackage{cite}
\usepackage{url}
\usepackage{enumitem}
\usepackage[usenames,dvipsnames]{xcolor}
\usepackage{todonotes}
\usepackage{multicol}
\usepackage{MnSymbol}
\usepackage{hyperref}
\usepackage{tikz-cd}
\usepackage[titletoc,title]{appendix}

\usetikzlibrary{arrows}

\AtBeginDocument{}%

\newcommand{\corurl}{red}
\newcommand{\corcite}{ForestGreen}
\newcommand{\corlink}{blue}

\hypersetup{linktocpage,colorlinks,urlcolor=\corurl,citecolor=\corcite,linkcolor=\corlink,
pdftitle={Functional evolution of scalar fields in one-dimensional bounded regions},
pdfauthor={J.F. Barbero, J. Margalef-Bentabol, E.J.S. Villaseñor}}

\numberwithin{equation}{section}  
\begin{document}


\bibliographystyle{plain}

\title{Functional evolution of scalar fields in bounded one-dimensional regions}

\author{
  {\small J. Fernando Barbero G.${}^{1,3}$, Juan Margalef-Bentabol${}^{1,2}$, and
          Eduardo J.S. Villase\~nor${}^{2,3}$} \\[4mm]
  {\small\it ${}^1$Instituto de Estructura de la Materia, CSIC} \\[-0.2cm]
  {\small\it Serrano 123, 28006 Madrid, Spain}         \\[1mm]
  {\small\it ${}^2$Grupo de Modelizaci\'on, Simulaci\'on Num\'erica
                   y Matem\'atica Industrial}  \\[-0.2cm]
  {\small\it Universidad Carlos III de Madrid} \\[-0.2cm]
  {\small\it Avda.\  de la Universidad 30, 28911 Legan\'es, Spain}            \\[1mm]
  {\small\it ${}^3$Grupo de Teor\'{\i}as de Campos y F\'{\i}sica
             Estad\'{\i}stica}\\[-2mm]
  {\small\it Instituto Gregorio Mill\'an, Universidad Carlos III de
             Madrid}\\[-2mm]
  {\small\it Unidad Asociada al Instituto de Estructura de la Materia, CSIC}
             \\[-2mm]
  {\small\it Madrid, Spain}           \\[-2mm]
  {\protect\makebox[5in]{\quad}}  
  \\
}
\date{November 29, 2016}
\maketitle
\thispagestyle{empty}   

\begin{abstract}
We discuss the unitarity of the quantum evolution between arbitrary Cauchy surfaces of a 1+1 dimensional free scalar field defined on a bounded spatial region and subject to several types of boundary conditions including Dirichlet, Neumann and Robin.
\end{abstract}

\medskip
\noindent
{\bf Key Words:}
arbitrary Cauchy surfaces; unitary quantum evolution; bounded domains; Bogoliubov coefficients.

\clearpage

%
%
\section{Introduction}{\label{sec_intro}}

The problem of the unitary implementation of the quantum dynamics of free fields under the evolution defined by sufficiently general, physically acceptable, spacetime slicings has received a lot of attention in the past. In \cite{Helfer} Helfer showed that, in general, the quantum dynamics cannot be unitarily implemented in curved spacetimes and hints that, even in the Minkowski spacetime, this may also happen when considering the evolution between arbitrary Cauchy surfaces.  Torre and Varadarajan \cite{TorreVar,TorreVar1,Torre} showed that this is indeed the case by studying the free scalar field on toroidal spatial slices (Cho and Varadarajan \cite{Var1} discussed Einstein-Rosen waves with a similar philosophy). The most important conclusion of these works is the realization of the fact that, beyond 1+1 dimensions, there are obstructions to the unitary implementation of the dynamics for certain types of seemingly good choices of Cauchy surfaces on a Minkowski spacetime. The authors of the quoted papers point out that, in fact, this is a \textit{generic} feature of the evolution of free quantum fields in more than one spatial dimension. An exception to this behavior is found for 1+1 dimensional models defined on the circle (the only closed one-dimensional manifold). In this case it is possible to show that the evolution can be unitarily implemented for arbitrary families of smooth spacelike slicings.

There are some issues that are not covered in the works that we have just quoted, for instance, the characterization of the families of embeddings capable of supporting unitary evolution in the $1+n$ dimensional case ($n>1$) or the consideration of spatial manifolds with boundary. The study of the latter problem is the main subject of the present paper. We hope that a straightforward extension of the methods that we use here will let us gain useful insights on the former.

Even in the simple one-dimensional setting, the introduction of boundaries ---and the consequent necessity of specifying boundary conditions to completely determine the classical dynamics-- introduces interesting changes regarding the unitary implementability of the quantum dynamics. This is so because, as we show in detail, the unitarity requirement forces us to work within particular classes of embeddings that cannot be arbitrarily chosen, in analogy with the generic situation in higher dimensions. The \textit{complete} characterization of the classes of spacelike embeddings capable of supporting unitary quantum evolution  of the free scalar field in a bounded one dimensional region is the main result of the paper.

From a technical point of view an important difference between our work and that of Torre and Varadarajan \cite{TorreVar,TorreVar1} lies in the fact that, if the spatial manifold is a circle, it suffices to consider evolution from a flat inertial hypersurface to an arbitrary (spacelike) one. This is so because the dynamics is always unitarily implementable in that case and, hence, between any two spacelike hypersurfaces. This is no longer true for the systems that we discuss in the paper, so we will need to consider the evolution between arbitrary hypersurfaces from the start.

Despite the fact that the final result, regarding the characterization of the suitable embeddings, is the same for all the boundary conditions, we show that there are interesting differences between Dirichlet and Neumann boundary conditions on one hand and Robin boundary conditions on the other. In fact, some of the difficulties of dealing with the higher dimensional cases are already present in the Robin case (which is quite interesting in its own right \cite{BMS}), hence, their satisfactory resolution suggests that a complete characterization of the embeddings capable of supporting unitary evolution for $n$-dimensional tori is possible.

We want to mention at this point that Agullo and Ashtekar \cite{AA} have recently proposed a novel approach to the study of unitarity in quantum field theories. The main ingredient of their approach is to use different quantizations for the different time slices. We do not follow this path here but our results are not incompatible with theirs. Rather, we suggest a possible way to \textit{select} spacetime slicings by demanding unitarity of the time evolution in the standard sense. It is important to point out that our results show that inertial slicings are always allowed but it is always possible to find more general spacetime slicings (containing any given Cauchy surface) that support unitary evolution.

The lay out of the paper is the following. After this introduction we study in section \ref{sec_Bogoliubov} the general form of the Bogoliubov coefficients for the evolution ---between two arbitrary Cauchy surfaces--- for a class of models incorporating the boundary conditions that we consider: Dirichlet, Neumann and Robin. We also give necessary and sufficient conditions to guarantee that the dynamics in each of these cases is unitary. Sections \ref{Dirichlet} and \ref{Robin} concentrate on the Dirichlet and Robin boundary conditions, respectively, and also discuss the Neumann boundary conditions as a particular subcase of the latter. We end the paper in section \ref{sec_conclusions} with our conclusions and some comments. A number of technical points on the embeddings that we use in the paper are discussed in appendix \ref{appendix_A}. The second appendix \ref{appendix_B} gives some information regarding the behaviour of some important quantities (i.e.\ eigenvalues of the Laplace operators, normalization coefficients and the like). Finally appendix \ref{appendix_C}  provides the proof of the main mathematical result employed in the characterization of the families of embeddings that can support unitary evolution.

%
%
\section{Bogoliubov coefficients and the unitary implementability of the quantum evolution}{\label{sec_Bogoliubov}}

Let us consider the classical evolution of a free, massless\footnote{The massive case is simpler to deal with. Its treatment is very similar to the one of the massless Robin case so we will not discuss it here. We work in the massless case to allow for zero modes.}, real scalar field $\varphi$ defined on the  manifold (with boundary) $\mathbb{R}\times [0,\pi]$ and subject to boundary conditions of Dirichlet, Neumann or Robin type. Here we are considering a spacetime (naturally embedded in $\mathbb{R}^2$) endowed with a Minkowskian metric $\eta$ of signature $(-,+)$ . When convenient we will use global coordinates $(t,x)$, denote the $t$ and $x$ derivatives as $\dot{\varphi}$ and $\varphi'$,  respectively, and write $\eta=-\mathrm{d}t^2+\mathrm{d}x^2$. The field equations are, simply,
\begin{equation}
\Box\varphi=0
\label{waveequation}
\end{equation}
where the fields are subject to boundary conditions at $x=0$ and $x=\pi$.

As it is well known, the space of solutions to \eqref{waveequation} can be endowed with a symplectic structure given by

\begin{equation}\label{simplectic}
\Omega(\varphi_1,\varphi_2)=\int_\Sigma\sqrt{\gamma_\Sigma}(\varphi_2{\L}_{n_\Sigma}\varphi_1-\varphi_1{\L}_{n_\Sigma}\varphi_2)\,,
\end{equation}
where $\varphi_1$ and $\varphi_2$ are two solutions to the field equations, $\Sigma$ is \textit{any} Cauchy hypersurface in $\mathbb{R}\times [0,\pi]$, ${\L}_{n_\Sigma}$ denotes the Lie derivative along the future directed unit normal to $\Sigma$ and $\gamma_\Sigma$ is the metric induced by $\eta$ on $\Sigma$. The metric volume form is $\sqrt{\gamma_\Sigma} \mathrm{d}\sigma$ where $\mathrm{d}\sigma$ is a fixed volume form on $\Sigma$ (that we will omit in many formulas). A straightforward argument that we sketch in the following shows that $\Omega$ is independent of the choice of $\Sigma$ for all the types of boundary conditions that we will employ in the paper. Indeed, in terms of the symplectic current
\begin{equation}\label{symplectic_current}
J(\varphi_1,\varphi_2):=\varphi_1\ast \mathrm{d}\varphi_2-\varphi_2\ast \mathrm{d}\varphi_1
\end{equation}
(with the Hodge dual defined as $\ast \mathrm{d}t=\mathrm{d}x$ and $\ast \mathrm{d}x=\mathrm{d}t$) we can write
\begin{equation}\label{symplectic_form}
  \Omega(\varphi_1,\varphi_2)=\int_\Sigma J(\varphi_1,\varphi_2)\,.
\end{equation}
As $\mathrm{d}J=0$, if we apply Stokes theorem and integrate $\mathrm{d}J$ on a region $R$ bounded by two non-intersecting spacelike hypersurfaces $\Sigma_1$, $\Sigma_2$ and $\partial:=R\cap(\{x=0\}\cup\{x=\pi\})$ we get
\begin{equation*}
  \int_{\Sigma_2}J- \int_{\Sigma_1}J+ \int_{\partial}J=0\,.
\end{equation*}
To conclude it suffices to check that $\imath_\partial^*J=(\varphi_1\varphi_2^\prime-\varphi_2\varphi_1^\prime)\mathrm{d}t=0$ for Dirichlet, Neumann and Robin boundary conditions and, hence, the integrals of $J$ on $\Sigma_1$ and $\Sigma_2$ are equal.

The solutions to the field equations \eqref{waveequation} can be expanded in terms of eigenfunctions of the different Laplacians
\begin{equation}\label{expansion_solutions}
  \varphi(t,x)=a_0(1-i t)Q_0(x)+a_0^*(1+i t)Q_0(x)+\sum_{k=1}^\infty (a_ke^{-i\omega_kt}+a_k^*e^{i\omega_kt})Q_k(x)\,,
\end{equation}
where the modes $Q_k(x)$ satisfy
\[Q_k^{\prime\prime}(x)=-\omega_k^2Q_k(x)\]
\textit{and the boundary conditions}. Here $Q_0$ denotes the zero mode that only appears in the Neumann case. In terms of the Fourier coefficients $a_k$ and $a^*_k$ the symplectic form reads
\begin{equation}\label{symplectic_form_modes}
  \Omega(\varphi_1,\varphi_2)=-i\sum_{k=1}^\infty 2\omega_kc_k^2(a_{1k}a^*_{2k}-a_{2k}a^*_{1k})-2c_0^2i(a_{10}a^*_{20}-a_{20}a^*_{10})\,.
\end{equation}
In order to have the standard expression for the symplectic form we choose the normalization constants $c_k$ as
\begin{equation*}
  c_0^2=\int_0^\pi Q_0^2(x)\mathrm{d}x=\frac{1}{2}\,,\quad c_k^2=\int_0^\pi Q_k^2(x)\mathrm{d}x=\frac{1}{2\omega_k}\,,
\end{equation*}
so that we have the orthogonality conditions
\begin{equation}\label{orthogonality}
  \int_0^\pi Q_{k_1}(x)Q_{k_2}(x)\mathrm{d}x=\frac{1}{2\omega_{k_1}}\delta_{k_1k_2}\,, \quad \int_0^\pi Q_0(x)Q_k(x)\mathrm{d}x=0\,,\quad k_1,k_2,k\in\mathbb{N}\,,
\end{equation}
and \eqref{symplectic_form_modes} is normalized.

In order to quantize and compute the Bogoliubov coefficients we need to know the scalar product in the space of complexified solutions of the field equations $\mathcal{S}_{\mathbb{C}}$ induced by the symplectic structure \eqref{simplectic}
\begin{equation}\label{Complexified_solution_space}
  \mathcal{S}_{\mathbb{C}}:=\{\varphi:\varphi=a_0^+\varphi_0^++\sum_{k=1}^\infty a_k^+\varphi_k^++a_0^-\varphi_0^-+\sum_{k=1}^\infty a_k^-\varphi_k^-\}
\end{equation}
where the positive ($s=+$) and negative ($s=-$) solutions are chosen as
\begin{equation*}
  \varphi_k^s(t,x):=e^{-i s \omega_k t}Q_k(x)\,,\quad \varphi_0^s:=(1-i s t)Q_0(x)\,.
\end{equation*}

The sesquilinear form
\begin{equation}\label{scalarproduct_1}
  \langle \varphi_1,\varphi_2\rangle:=-i\int_\Sigma\sqrt{\gamma_\Sigma}(\varphi_2{\L}_{n_\Sigma}\varphi_1^*-\varphi_1^*{\L}_{n_\Sigma}\varphi_2)
\end{equation}
defines a (positive definite) scalar product on the positive frequency subspace of $\mathcal{S}_{\mathbb{C}}$. In an arbitrary Cauchy hypersurface given by $\mathbf{X}(x)=(T(x),X(x))$ in inertial coordinates (see appendix \ref{appendix_A}), the expression for \eqref{scalarproduct_1} is
\begin{align}\label{scalarproduct_2}
  \langle \varphi_{k_1}^{s_1},\varphi_{k_2}^{s_2}\rangle & =\int_0^\pi |\gamma_{\mathbf{X}}|^{1/2} e^{-i(s_2\omega_{k_2}-s_1\omega_{k_1})T}(s_2 \omega_{k_2}+s_1 \omega_{k_1})n_{\mathbf{X}}^0Q_{k_1}(X)Q_{k_2}(X) \\
   - i \int_0^\pi & |\gamma_{\mathbf{X}}|^{1/2} e^{-i(s_2\omega_{k_2}-s_1\omega_{k_1})T}(s_2 \omega_{k_2}+s_1 \omega_{k_1})n_{\mathbf{X}}^1\big(Q_{k_1}^\prime(X)Q_{k_2}(X) -Q_{k_1}(X)Q_{k_2}^\prime(X)\big)\,,\nonumber
\end{align}
where we denote the unit normal to the hypersurface $\mathbf{X}(\Sigma)$ as $n_{\mathbf{X}}(x)=(n_{\mathbf{X}}^0(x),n_{\mathbf{X}}^1(x))$, and similar formulas apply when zero modes are present (i.e.\ for Neumann boundary conditions).

In terms of the modes $Q_k$, and taking into account that in $1+1$ dimensions the unit future pointing normal $n$ is given by
\[
n=(n^0,n^1)=\frac{1}{\sqrt{\gamma_{\mathbf{X}}}}(X',T')\,,
\]
the Bogoliubov coefficients for the field evolution between two spatial hypersurfaces $\mathbf{X}_1(\Sigma)$ and $\mathbf{X}_2(\Sigma)$ can be written in the form
\begin{align}\label{Bogo_coeffs}
\beta_{k_1 k_2}^{\bm{X}_1\bm{X}_2} &= \int_\Sigma \big(\omega_{k_1}X_1'Q_{k_1}(X_1)Q_{k_2}(X_2)-iT_1'Q'_{k_1}(X_1)Q_{k_2}(X_2) \big) e^{i\omega_{k_2}T_2+i\omega_{k_1}T_1} \\
   & -\int_\Sigma  \big(\omega_{k_2}X'_2Q_{k_1}(X_1)Q_{k_2}(X_2)-iT_2'Q_{k_1}(X_1)Q'_{k_2}(X_2) \big) e^{i\omega_{k_2}T_2+i\omega_{k_1}T_1}\,.\nonumber
\end{align}
Following \cite{Shale}, the necessary and sufficient condition that guarantees the unitary implementability of the quantum evolution given by the Cauchy slices associated with $\bm{X}_1$ to $\bm{X}_2$ is
\[\sum_{k_1,k_2}|\beta_{k_1 k_2}^{\bm{X}_1\bm{X}_2}|^2<\infty\,.\]
In the following sections we study when this condition is satisfied and characterize the families of embeddings that support unitary evolution for the different types of boundary conditions used in the paper.

%
%
\section{Dirichlet boundary conditions}\label{Dirichlet}

The normalized eigenfunctions of the Laplace operator with Dirichlet boundary conditions in the interval $[0,\pi]$ are
\begin{equation}\label{eigen_Dirichlet}
  Q_k(x)=\frac{1}{\sqrt{\pi k}}\sin kx\,,\quad \omega_k=k\,,\quad k\in \mathbb{N}\,.
\end{equation}
There are no zero modes in this case. By introducing these in \eqref{Bogo_coeffs} we immediately obtain
\begin{align*}
\beta_{k_1 k_2}^{\bm{X}_1\bm{X}_2}
&= \frac{1}{4\pi \sqrt{k_1k_2}}\int_0^\pi (k_2X'_2-k_1X_1'-k_1T_1'+k_2T'_2) e^{i (k_1(T_1+X_1)+k_2(T_2+X_2))}\, \mathrm{d}\sigma \\
&+ \frac{1}{4\pi \sqrt{k_1k_2}}\int_0^\pi (k_2X'_2-k_1X_1'+k_1T_1'-k_2T'_2) e^{i (k_1(T_1-X_1)+k_2(T_2-X_2))}\, \mathrm{d}\sigma\\
&+\frac{1}{4\pi \sqrt{k_1k_2}}\int_0^\pi (k_1X'_1-k_2X_2'+k_1T_1'+k_2T'_2) e^{i (k_1(T_1+X_1)+k_2(T_2-X_2))}\, \mathrm{d}\sigma \\
&+\frac{1}{4\pi \sqrt{k_1k_2}}\int_0^\pi (k_1X'_1-k_2X_2'-k_1T_1'-k_2T'_2) e^{i (k_1(T_1-X_1)+k_2(T_2+X_2))}\, \mathrm{d}\sigma\,.
\end{align*}
In order to compute the first two integrals in the preceding expression we closely follow the procedure introduced by Torre and Varadarajan (see appendix in \cite{TorreVar}). For instance, for the first integral $I_1$ we use the change of variables $u=\lambda(T_1+X_1)+(1-\lambda)(T_2+X_2)$ and integrate by parts [$\lambda\in(0,1)$ is defined as $\lambda=\lambda(k_1,k_2):=k_1/(k_1+k_2)\,$]. This gives
\begin{align*}
&I_1=\frac{i}{4\pi\sqrt{k_1k_2}(k_1+k_2)}\left(e^{i(k_1T_1(0)+k_2T_2(0))}\frac{k_2(X_2^\prime(0)+T_2^\prime(0))-k_1(X_1^\prime(0)+T_1^\prime(0))}{\lambda (T_1^\prime(0)+X_1^\prime(0))+(1-\lambda)(T_2^\prime(0)+X_2^\prime(0))}\right.\\
&\left.-(-1)^{k_1+k_2}e^{i(k_1T_1(\pi)+k_2T_2(\pi))}\frac{k_2(X_2^\prime(\pi)+T_2^\prime(\pi))-k_1(X_1^\prime(\pi)+
T_1^\prime(\pi))}{\lambda(T_1^\prime(\pi)+X_1^\prime(\pi))+(1-\lambda)(T_2^\prime(\pi)+X_2^\prime(\pi))}+O\left(\frac{1}{k_1+k_2}\right)\right)\,.
\end{align*}
A completely analogous procedure can be used to compute the second integral. Finally the last two integrals can be trivially obtained in closed form. Putting all this together we find that the Bogoliubov coefficients in this case are
\begin{align}
\hspace*{-4mm}\beta_{k_1 k_2}^{\bm{X}_1\bm{X}_2}
= &\frac{i\sqrt{k_1k_2}}{\pi (k_1+k_2)^2} \left(
\frac{(-1)^{k_1+k_2}e^{i(k_1T_1(\pi)+k_2T_2(\pi))}(X_1'(\pi)T_2'(\pi)-T_1'(\pi)X_2'(\pi))}{(\lambda T'_1(\pi)+(1-\lambda)T'_2(\pi))^2-(\lambda X'_1(\pi)+(1-\lambda)X'_2(\pi))^2}
   \right.\label{Bogo_Dirichlet}\\
   &\left.
   -\frac{e^{i(k_1T_1(0)+k_2T_2(0))}(X_1'(0)T_2'(0)-T_1'(0)X_2'(0))}{(\lambda T'_1(0)+(1-\lambda)T'_2(0))^2-(\lambda X'_1(0)+(1-\lambda)X'_2(0))^2}+O\left(\frac{1}{k_1k_2}\right)
   \right)\,.\nonumber
\end{align}
The Bogoliubov coefficients for Dirichlet boundary conditions \eqref{Bogo_Dirichlet} have the form
\begin{equation}\label{Bogo-coeffs_AB}
\hspace*{-10mm}\beta_{k_1 k_2}^{\bm{X}_1\bm{X}_2}=\frac{\sqrt{k_1 k_2}}{\pi(k_1+k_2)^2}\left(\gamma^{\bm{X}_1\bm{X}_2}_{k_1k_2}+O\left(\frac{1}{k_1k_2}\right)\right)\,,
\end{equation}
where
\begin{equation}\label{gamma_Dirich}
\gamma^{\bm{X}_1\bm{X}_2}_{k_1k_2}:=i e^{i(k_1\alpha_{1\sigma}+k_2\alpha_{2\sigma})}f_\sigma(\lambda) V_\sigma\Big|_{\sigma=0}^{\sigma=\pi}
\end{equation}

\begin{align}
  V_\sigma:=X_1'(\sigma)T_2'(\sigma) -T_1'(\sigma)X_2'(\sigma)\,,\label{Vsigma}
\end{align}
$\alpha_{i0}$ and $\alpha_{i\pi}$ ($i=1,\,2$) are constants defined in terms of the end-points of the embeddings
\begin{align}
  \alpha_{10}:=&T_1(0)\,, \,\,\,\,\,\quad\quad \alpha_{20}:=T_2(0) \label{betas}\\
  \alpha_{1\pi}:=&T_1(\pi)+\pi\,, \quad \alpha_{2\pi}:=T_2(\pi)+\pi \label{alfas}
\end{align}
and
\[
\lambda=\lambda(k_1,k_2):=\frac{k_1}{k_1+k_2}\in (0,1)
\]
and the functions $f_0,\,f_\pi:[0,1]\rightarrow \mathbb{R}$ are given by
\begin{align}
  f_\sigma(\lambda) &=\frac{1}{(\lambda T_1'(\sigma)+(1-\lambda)T_2'(\sigma))^2-(\lambda X_1'(\sigma)+(1-\lambda)X_2'(\sigma))^2}\,.\label{f0}
\end{align}

As we show in appendix \ref{appendix_C} the necessary and sufficient condition to guarantee the convergence of $\sum_{k_1,k_2}|\beta_{k_1 k_2}^{\bm{X}_1\bm{X}_2}|^2$ is the vanishing of the coefficients $V_0$ and $V_\pi$ defined in \eqref{Vsigma}. This implies that
\begin{equation}\label{conditions_Dirichlet}
  \frac{T_1'(\pi)}{X_1'(\pi)}=  \frac{T_2'(\pi)}{X_2'(\pi)}\,,\quad  \frac{T_1'(0)}{X_1'(0)}=  \frac{T_2'(0)}{X_2'(0)}\,.
\end{equation}
The interpretation of these conditions is straightforward. Unitary evolution requires that the slope of the embedded surfaces both at $0$ and $\pi$ must be separately preserved under the evolution of the system. Notice, by the way, that for any embedding  $X'(0)\neq0$ and $X'(\pi)\neq0$ (see appendix \ref{appendix_A}).

It is interesting to notice that (for non-zero modes) in the Neumann case the Bogoliubov coefficients differ only in a global sign. In this case
\[
 Q_k(x)=\frac{1}{\sqrt{\pi k}}\cos kx\,,\quad  \omega_k=k\,,\quad k\in\mathbb{N}\,,
\]
and hence  $\beta_{k_1 k_2}^{\bm{X}_1\bm{X}_2}$ for $k_1\,,k_2\in \mathbb{N}$ are simply \eqref{Bogo_Dirichlet} multiplied by $-1$. It is straightforward to see that $\beta_{0 k_2}^{\bm{X}_1\bm{X}_2}=\beta_{k_10}^{\bm{X}_1\bm{X}_2}=0$ and the concrete value of $\beta_{00}^{\bm{X}_1\bm{X}_2}$ is irrelevant. From this we conclude that the Neumann case essentially reduces to the Dirichlet case as far as the unitarity of the evolution is concerned.

%
%
\section{Robin boundary conditions}\label{Robin}

Let us consider now the problem with Robin boundary conditions of the form
\begin{align}
&Q_k^\prime(0)-\kappa_0 Q_k(0)=0 \label{Robin0}\\
&Q_k^\prime(\pi)+\kappa_\pi Q_k(\pi)=0\label{RobinPi}
\end{align}
with $k\in\mathbb{N}$ and $\kappa_0,\kappa_\pi\geq 0$. There are zero modes if and only if $\kappa_0=\kappa_\pi=0$ (Neumann boundary conditions) so we will concentrate in the following in modes with $\omega>0$. Mixed Robin-Dirichlet boundary conditions can be treated in a straightforward way so we will not discuss them here.

The eigenfunctions of the Laplace operator subject to these boundary conditions satisfy the equation $Q_k^{\prime\prime}=-\omega_k^2Q_k$ with $\omega_k$ given by the positive solutions to the equation
\begin{equation}
\label{ec_robin}
(\omega_k^2-\kappa_0\kappa_\pi)\sin\pi\omega_k-(\kappa_0+\kappa_\pi)\omega_k\cos\pi\omega_k=0\,.
\end{equation}
These eigenfunctions can be written as a linear combination of $\sin\omega_k x$ and $\cos\omega_k x$. However, in order to keep the analogy with the Dirichlet case it is convenient to write them in the form
\begin{equation}
Q_k(x)=C_k\sin(\omega_k x+\phi_k)\,,\label{eigenfunctions_Robin}
\end{equation}
where we have absorbed the constants multiplying $\sin\omega_k x$ and $\cos\omega_k x$ in a phase $\phi_k\in (0,\pi/2]$ and an amplitude $C_k>0$ satisfying
\begin{align*}
&\sin \phi_k=\frac{\omega_k}{\sqrt{\omega_k^2+\kappa_0^2}}\,,\quad \cos \phi_k=\frac{\kappa_0}{\sqrt{\omega_k^2+\kappa_0^2}}\,,\quad \phi_k=\arctan\frac{\omega_k}{\kappa_0}\,,\\
& \frac{1}{C^2_k}=\pi\omega_k\left( 1+\frac{\kappa_0+\kappa_\pi}{\pi}\frac{\omega_k^2+\kappa_0\kappa_\pi}{(\omega_k^2+\kappa_0^2)(\omega_k^2+\kappa_{\pi}^2)}\right)\,.
\end{align*}
The Robin boundary conditions \eqref{Robin0}-\eqref{RobinPi} imply that
\begin{align*}
&\omega_k\cos\phi_k=\kappa_0\sin\phi_k\,,\\
&\omega_k\cos(\pi\omega_k+\phi_k)=-\kappa_\pi\sin(\pi\omega_k+\phi_k)\,,
\end{align*}
and, hence, we can derive the following identities (valid for $\sigma=0,\pi$):
\begin{align}
\hspace*{-0.5cm}\kappa_\sigma\sin(\omega_{k_1}\sigma+\omega_{k_2}\sigma+\phi_{k_1}+\phi_{k_2})=&(-1)^{\frac{\sigma}{\pi}}(\omega_{k_1}+\omega_{k_2})\cos(\omega_{k_1}\sigma+\phi_{k_1})\cos(\omega_{k_2}\sigma+\phi_{k_2})\,,\label{identidad1}\\
\hspace*{-0.5cm}\kappa_\sigma\sin(\omega_{k_1}\sigma-\omega_{k_2}\sigma+\phi_{k_1}-\phi_{k_2})=&(-1)^{\frac{\sigma}{\pi}}(\omega_{k_1}-\omega_{k_2})\cos(\omega_{k_1}\sigma+\phi_{k_1})\cos(\omega_{k_2}\sigma+\phi_{k_2})\,.\label{identidad2}
\end{align}
Using the form of the eigenfunctions \eqref{eigenfunctions_Robin}, the general expression for the Bogoliubov coefficients \eqref{Bogo_coeffs} and following the same steps as in the Dirichlet case we get
\begin{align*}
\beta_{k_1 k_2}^{\bm{X}_1\bm{X}_2}
&= \frac{C_{k_1}C_{k_2}}{4}\int_0^\pi (\omega_{k_2}X'_2-\omega_{k_1}X_1'-\omega_{k_1}T_1'+\omega_{k_2}T'_2) e^{i (\omega_{k_1}(T_1+X_1)+\omega_{k_2}(T_2+X_2))+\phi_{k_1}+\phi_{k_2})}\, \mathrm{d}\sigma \\
&+ \frac{C_{k_1}C_{k_2}}{4}\int_0^\pi (\omega_{k_2}X'_2-\omega_{k_1}X_1'+\omega_{k_1}T_1'-\omega_{k_2}T'_2) e^{i (\omega_{k_1}(T_1-X_1)+\omega_{k_2}(T_2-X_2))-\phi_{k_1}-\phi_{k_2})}\, \mathrm{d}\sigma\\
&+\frac{C_{k_1}C_{k_2}}{4}\int_0^\pi (\omega_{k_1}X'_1-\omega_{k_2}X_2'+\omega_{k_1}T_1'+\omega_{k_2}T'_2) e^{i (\omega_{k_1}(T_1+X_1)+\omega_{k_2}(T_2-X_2))+\phi_{k_1}-\phi_{k_2})}\, \mathrm{d}\sigma \\
&+ \frac{C_{k_1}C_{k_2}}{4}\int_0^\pi (\omega_{k_1}X'_1-\omega_{k_2}X_2'-\omega_{k_1}T_1'-\omega_{k_2}T'_2) e^{i (\omega_{k_1}(T_1-X_1)+\omega_{k_2}(T_2+X_2))-\phi_{k_1}+\phi_{k_2})}\, \mathrm{d}\sigma\,.
\end{align*}
As we can see the structure of these integrals is very similar to those appearing in the computation of the Bogoliubov coefficients for the Dirichlet case; they can be obtained along the same lines. They are
\begin{align}
\beta_{k_1 k_2}^{\bm{X}_1\bm{X}_2}
&=\frac{\sqrt{\omega_{k_1}\omega_{k_2}}}{\pi(\omega_{k_1}+\omega_{k_2})^2}\left(\gamma^{\bm{X}_1\bm{X}_2}_{k_1k_2}+O\left(\frac{1}{\omega_{k_1}\omega_{k_2}}\right)\right)\label{Bogo_Robin}
\end{align}
where
\begin{align}
\gamma^{\bm{X}_1\bm{X}_2}_{k_1 k_2}:=\pi\sqrt{\omega_{k_1}\omega_{k_2}}C_{k_1}C_{k_2}e^{i\psi_\sigma(k_1,k_2)}f_\sigma(\tau)\Big( &
iV_\sigma\cos(\omega_{k_1}\sigma+\phi_{k_1}+\omega_{k_2}\sigma+\phi_{k_2})
\label{gamma_Robin}\\&\hspace*{-2.3cm}+
\kappa_\sigma g_\sigma(\tau)\frac{\omega_{k_1}+\omega_{k_2}}{\omega_{k_1}\omega_{k_2}}\sin(\omega_{k_1}\sigma+\phi_{k_1}) \sin(\omega_{k_2}\sigma+\phi_{k_2})\Big)\Big|_{\sigma=0}^{\sigma=\pi}\,.
\nonumber
\end{align}
In analogy with the Dirichlet case we have introduced the notation
\[\tau=\tau(\omega_{k_1},\omega_{k_2}):=\frac{\omega_{k_1}}{\omega_{k_1}+\omega_{k_2}}\in (0,1)\,,\]
The expressions for $f_\sigma$ and $V_\sigma$ are given in \eqref{f0} and \eqref{Vsigma}. We have also introduced
\begin{align*}
&g_\sigma(\tau):=\tau (N_1^2(\sigma)-S(\sigma))-(1-\tau)(N_2^2(\sigma)-S(\sigma))\\
&\psi_\sigma(k_1,k_2):=\omega_{k_1}T_1(\sigma)+\omega_{k_2}T_2(\sigma)\,,
\end{align*}
and
\begin{align*}
&N^2_i(\sigma):=T_i^{\prime 2}(\sigma)-X^{\prime2}(\sigma)\,, i=1,2\\
&S(\sigma):=T_1'(\sigma)T_2'(\sigma)-X_1'(\sigma)X_2'(\sigma)\,.
\end{align*}
Notice that, modulo a global sign, \eqref{gamma_Robin} reduces to \eqref{gamma_Dirich} in the Neumann case ($\kappa_0=\kappa_\pi=0$). In order to see this, the equations appearing in appendix \ref{appendix_B} are useful.

The discussion of the convergence of the series
\[
\sum_{k_1,k_2}|\beta_{k_1 k_2}^{\bm{X}_1\bm{X}_2}|^2
\]
is lengthy but straightforward. We just point out the most important features of the analysis.

The modulus squared of the Bogoliubov coefficients $|\beta_{k_1 k_2}^{\bm{X}_1\bm{X}_2}|^2$ consists of three types of terms: those involving $V_\sigma^2$, those with $g_\sigma^2$ and terms with phases $e^{\pm i(\psi_\pi-\psi_0)}$. The terms proportional to $V_\sigma^2$, as in the Dirichlet case, diverge; those with $g_\sigma^2$ converge; and an argument that relies on the result proved in appendix \ref{appendix_C} shows that the terms involving the phases also converge. As a consequence of this, the necessary and sufficient condition guaranteing the unitarity of the dynamics is, as in the Dirichlet case, the vanishing of $V_\sigma$.

%
%
\section{Conclusions and comments}{\label{sec_conclusions}}

We have analyzed the dynamical evolution  of a quantum free scalar field defined on a flat spacetime of the form $\mathbb{R}\times [0,\pi]$ satisfying Dirichlet, Neumann or Robin boundary conditions. Along the way we have characterized the equivalence classes of spacelike embeddings that support unitary evolution. These classes are labeled by the values of $T'(0)/X'(0)$ and $T'(\pi)/X'(\pi)$, in the sense that for any two embeddings $\bm{X}_1$, $\bm{X}_2$ satisfying $T_1'(0)/X_1'(0)=T_2'(0)/X_2'(0)$ and $T_1'(\pi)/X_1'(\pi)=T_2'(\pi)/X_2'(\pi)$, the field dynamics between $\bm{X}_1(\Sigma)$ and $\bm{X}_2(\Sigma)$ can be unitarily implemented. Notice that the simultaneity hypersurfaces defined by an inertial (free) observer always define a slicing of spacetime for which the dynamics can be unitarily implemented. The Schr\"odinger picture of the functional evolution is not available in general (as in the  $1+n$ dimensional case) but it is well defined within the equivalence class of embeddings just mentioned.

Several comments are in order now: The main reason to remain within the usual Fock quantization ---as we do here--- is the ease to physically interpret the states as vectors in a Hilbert space with simple properties. It is in this framework that the phenomenon of the impossibility to unitarily implement quantum evolution shows up. As emphasized by a number of authors, there are other possible approaches to the quantization of the field models that we discuss in this paper that can be used to circumvent  some of the problems associated with the lack of unitary evolution in a satisfactory way. For instance, in the algebraic approach to quantum field theory (see for example \cite{Wald} and the discussion in \cite{TorreVar1,Torre}), the $C^*$ algebra $\mathcal{A}$ of basic observables of a free theory is taken to be the Weyl algebra. $\mathcal{A}$ is generated by elements $W(\varphi)$, labeled
by points $\varphi$ of the covariant phase space (i.e.\ the space of solutions to the field equations), satisfying
\[
W(\varphi)^*= W(-\varphi)\,,\quad W(\varphi_1)W(\varphi_2)= e^{-i\Omega(\varphi_1,\varphi_2)}W(\varphi_1 + \varphi_2). \]
The states are  positive, normalized, linear functions $\omega:\mathcal{A} \rightarrow  \mathbb{C}$.  In this approach \cite{TorreVar1}, if we consider two Cauchy hypersurfaces represented by the embeddings $\bm{X}_1$ and $\bm{X}_2$ and we assign the state $\omega$ to the Cauchy hypersurface $\bm{X}_1(\Sigma)$, the expectation value of the observable represented by element $W(\varphi)$ of the Weyl algebra at the Cauchy hypersurface $\bm{X}_2(\Sigma)$ is always well defined and is given by
\[
\langle W(\varphi)\rangle_{(\omega,\bm{X}_1, \bm{X}_2)} = \omega(W(\mathcal{T}^{-1}_{(\bm{X}_1,\bm{X}_2)}\varphi))\,.
\]
where $\mathcal{T}_{(\bm{X}_1,\bm{X}_2)}$ is a bijection on the solution space $\Gamma$ defined in terms of the map $\mathcal{I}_{\bm{X}}$ that associates a solution to the field equations to particular initial data given on the Cauchy surface $\bm{X}(\Sigma)$. Notice that on the Cauchy hypersurface $\bm{X}_1(\Sigma)$ the expectation value is given by
\[\langle W(\varphi)\rangle_{(\omega,\bm{X}_1,\bm{X}_1)} = \omega(W(\varphi)).\]
No unitarity problem arises in this context in the Heisenberg picture, as long as one is only interested in the evolution of observables in the Weyl algebra. Of course, in order to study the physics associated with observables that do not belong to this algebra it is necessary to use other methods.

A possible solution to this issue (partially anticipated by Torre and Varadarajan in \cite{TorreVar1}) has been suggested by Agullo and Ashtekar in \cite{AA}. Given a foliation $\bm{X}_t$, $t\in \mathbb{R}$,  one can construct a $1$-parameter family of representations $\bm{R}_t$ on Hilbert spaces $(\mathcal{H}_t,\langle\cdot,\cdot\rangle_t)$ of the Weyl algebra $\mathcal{A}$ in such a way that \textit{there exists} a family of unitary operators $U(t_2,t_1): \mathcal{H}_{t_1}\rightarrow  \mathcal{H}_{t_2}$ such that
\[
\langle \Phi_1 , \bm{R}_{t_1}(W(\mathcal{T}^{-1}_{(\bm{X}_{t_1},\bm{X}_{t_2})}\varphi)) \Psi_1\rangle_{t_1}=\langle U(t_2,t_1)\Phi_1 , \bm{R}_{t_2}(W(\varphi)) U(t_2,t_1)\Psi_1\rangle_{t_2}\,.
\]
In this approach evolution is always unitary by construction.

As we have shown in the paper, the presence of boundaries prevents us from reaching the conclusions of \cite{TorreVar1}. The main reason is that the argument used in that paper to show that the dynamics can be unitarily implemented relies on the possibility of performing successive integrations by parts involving smooth functions to show that the Bogoliubov coefficients decay sufficiently fast. In the examples that we have considered here the surface terms appearing after integration by parts spoil the simple argument of \cite{TorreVar1} and require a careful discussion.

The analysis that we have presented here is more complicated than the one of \cite{TorreVar1} because we are forced to consider the evolution between \textit{arbitrary} Cauchy hypersurfaces (and not just from an inertial one to an arbitrary one) as the dynamics between generic Cauchy hypersurfaces cannot be unitarily implemented.


An interesting application of our results is the polymer quantization of this class of models generalizing the results already obtained in \cite{Var2,LV1,LV2,LV3} for the circle. This may help understand the quantization of diff-invariant field theories in the presence of boundaries and may find applications in the study of black holes in loop quantum gravity \cite{BV} (where they are modelled with the help of spacetime boundaries called \textit{isolated horizons} \cite{AK}).

%
%
\section*{Acknowledgments}

 The authors wish to thank Ivan Agullo and Madhavan Varadarajan for their valuable comments. This work has been supported by the Spanish MINECO research grant FIS2014-57387-C3-3-P. Juan Margalef-Bentabol is supported by a ``la Caixa'' fellowship and a Residencia de Estudiantes (MINECO) fellowship.

\begin{appendices}

%
%
\section{Some facts about the spatial embeddings}\label{appendix_A}

The spacelike embeddings $\mathbf{X}:=(T,X)$ that we use in the paper can be defined in terms of a pair of $C^\infty([0,\pi])$ functions\footnote{This guarantees, among other things, that the integrals that are left after the several integration by parts that we perform in the paper are well defined.} $T,X:[0,\pi]\rightarrow \mathbb{R}$ satisfying the conditions $X(0)=0$, $X(\pi)=\pi$. The fact that they are required to be spacelike means that
\begin{equation}\label{spacelike}
-T'(s)^2+X'(s)^2>0
\end{equation}
for all $s\in [0,\pi]$. As a consequence of this (and the required orientability) we have $X'(s)>0$ for all $s\in[0,\pi]$ and, hence, \eqref{spacelike} is equivalent to either of the following:
\begin{align}
\left|\frac{T'(s)}{X'(s)}\right|<1\,, &\quad\forall s\in[0,\pi]\,,\label{spacelike1}\\
-X'(s)<T'(s)<X'(s) \,, &\quad\forall s\in[0,\pi]\,.\label{spacelike2}
\end{align}
Integrating \eqref{spacelike2} on the interval $[0,\pi]$ and using $X(0)=0$, $X(\pi)=\pi$ it is straightforward to show that $|T(\pi)-T(0)|<\pi$ which implies that $\gamma_i:=T_i(\pi)-T_i(0)+\pi$, ($i=1,2$ and $\mathbf{X}_i$ are two spacelike embeddings) satisfy $0<|\gamma_i|<2\pi$, conditions that play a relevant role in the analysis presented in appendix \ref{appendix_C}.

If we have two spacelike embeddings $\mathbf{X}_i$, $i=1,2$ and take $\lambda\in(0,1)$, the conditions \eqref{spacelike2} imply
\begin{align*}
   & -\lambda X_1'(\pi)<\lambda T_1'(\pi)< \lambda X_1'(\pi)\\
   & -(1-\lambda) X_2'(\pi)<(1-\lambda)T_2'(\pi)<(1-\lambda)X_2'(\pi)\,.
\end{align*}
Adding them up we see that $|\lambda T_1'(\pi)+(1-\lambda)T_2'(\pi)|<|\lambda X_1'(\pi)+(1-\lambda)X_2'(\pi)|$. Taking into account that $T_i'(\pi)^2-X_i'(\pi)^2<0$, for $i=1,2$ we conclude that for all $\lambda\in[0,1]$ we have
\[
(\lambda T_1'(\pi)+(1-\lambda)T_2'(\pi))^2-(\lambda X_1'(\pi)+(1-\lambda)X_2'(\pi))^2<0\,.
\]
In the same way we can prove that
\[
(\lambda T_1'(0)+(1-\lambda)T_2'(0))^2-(\lambda X_1'(0)+(1-\lambda)X_2'(0))^2<0\,.
\]
With the help of these last two conditions it is straightforward to show that the functions $f_0$ and $f_\pi$ defined by (\ref{f0}) can be extended to the closed interval $[0,1]$ in such a way that they are infinitely differentiable, i.e.\ they are bounded in $[0,1]$ together with all their derivatives. The same conclusion applies to $h:=f_0f_\pi$. In the following the extensions of these functions will be denoted with the same root letter.

%
%
\section{Some details about the Robin case}\label{appendix_B}

In this appendix we give some complementary information about the eigenvalues and eigenfunctions \eqref{eigenfunctions_Robin} for Robin boundary conditions.

Asymptotically $\omega_k$ behaves as
\begin{equation}\label{asympt_omega}
\omega_{k+1}=k+\frac{\kappa_0+\kappa_\pi}{\pi k}-\left(\frac{(\kappa_0+\kappa_\pi)^2}{\pi^2}+\frac{\kappa_0^3+\kappa_\pi^3}{3\pi}\right)\frac{1}{k^3}+O\left(\frac{1}{k^5}\right)\,,\quad k\rightarrow\infty\,.
\end{equation}
when $k\rightarrow\infty$. It is also useful to know the asymptotic behavior in the same limit of $\phi_k$
\begin{equation}
\phi_{k+1}=\frac{\pi}{2}-\frac{\kappa_0}{k}+\frac{3\kappa_0^2+\kappa_0^3+3\kappa_0\kappa_\pi}{3k^3}+O\left(\frac{1}{k^5}\right)\,,\label{asympt_phi}
\end{equation}
and
\begin{equation}
C_{k+1}\sqrt{\pi\omega_{k+1}}=1-\frac{\kappa_0+\kappa_\pi}{2\pi k^2}+O\left(\frac{1}{k^4}\right)\,.\label{asympt_Rk}
\end{equation}
The following identities are also useful
\begin{align*}
   & \cos(\pi \omega_{k_1}+\phi_{k_1}+\pi \omega_{k_2}+\phi_{k_2})=(-1)^{k_1+k_2}\frac{\kappa_\pi^2-\omega_{k_1}\omega_{k_2}}{\sqrt{(\omega_{k_1}^2+\kappa_\pi^2)(\omega_{k_2}^2+\kappa_\pi^2)}}\,, \\
   & \cos(\phi_{k_1}+\phi_{k_2})=\frac{\kappa_0^2-\omega_{k_1}\omega_{k_2}}{\sqrt{(\omega_{k_1}^2+\kappa_0^2)(\omega_{k_2}^2+\kappa_0^2)}}\,,\\
   &\sin(\pi\omega_k+\phi_k)=(-1)^k\frac{\omega_k}{\sqrt{\omega_k^2+\kappa_\pi^2}}\,,\\
   &\cos(\pi\omega_k+\phi_k)=(-1)^k\frac{\kappa_\pi}{\sqrt{\omega_k^2+\kappa_\pi^2}}\,,\\
   &\sin(\pi\omega_k)=(-1)^{k+1}\frac{(\kappa_0+\kappa_\pi)\omega_k}{\sqrt{(\omega_k^2+\kappa_0^2)(\omega_k^2+\kappa_\pi^2)}}\,,\\
   &\cos(\pi\omega_k)=(-1)^{k+1}\frac{\omega_k^2-\kappa_0\kappa_\pi}{\sqrt{(\omega_k^2+\kappa_0^2)(\omega_k^2+\kappa_\pi^2)}}\,.
\end{align*}

%
%
\section{Series convergence of the Bogoliubov coefficients for Dirichlet and Neumann boundary conditions}\label{appendix_C}

Let us prove that the vanishing of $V_0$ and $V_\pi$, defined in (\ref{Vsigma}), are the necessary and sufficient conditions for the convergence of
\[\sum_{k_1,k_2}|\beta_{k_1 k_2}^{\bm{X}_1\bm{X}_2}|^2<\infty\,,\quad k_1,\,k_2\in\mathbb{N}\]
if the Bogoliubov coefficients have the form \eqref{Bogo-coeffs_AB}:
\[
\beta_{k_1 k_2}^{\bm{X}_1\bm{X}_2}=\frac{\sqrt{k_1 k_2}}{\pi(k_1+k_2)^2}\left(V_\pi e^{i(k_1\alpha_{1\pi}+k_2\alpha_{2\pi})}f_\pi(\lambda)-V_0e^{i(k_1\alpha_{10}+k_2\alpha_{20})}f_0(\lambda)+O\left(\frac{1}{k_1+k_2}\right)\right)\,.
\]
The only non-trivial part of the proof is showing that the conditions are necessary. To this end we need to consider the following double series given by the sum of the squares of the absolute values of the ``leading part'' of the Bogoliubov coefficients
\[
\sum_{k_1,k_2\in\mathbb{N}}\frac{k_1k_2}{(k_1+k_2)^4}\Big(V_\pi^2f_\pi^2(\lambda)+V_0^2f_0^2(\lambda)-2V_0V_\pi f_0(\lambda)f_\pi(\lambda)\mathrm{Re}\left(e^{ik_1(\alpha_{1\pi}-\alpha_{10})+ik_2(\alpha_{2\pi}-\alpha_{20})}\right)\Big)\,.
\]
As this is a series of positive terms it will converge if and only if the following ordinary (``simple'') series does
\begin{equation}\label{seriesimple}
\hspace*{-5mm}\sum_{j=2}^\infty \sum_{k=1}^{j-1}\frac{k(j-k)}{j^4}\left(V_\pi^2f_\pi^2\left(\frac{k}{j}\right)+V_0^2f_0^2\left(\frac{k}{j}\right)-2V_0V_\pi f_0\left(\frac{k}{j}\right)f_\pi\left(\frac{k}{j}\right)\mathrm{Re}\left(e^{ik(\gamma_1-\gamma_2)+i\gamma_2j}\right)\right).
\end{equation}
As we have shown in appendix \ref{appendix_A} the condition that the embeddings are spacelike implies that $0<|\gamma_1|,\, |\gamma_2|<2\pi$.

The gist of the argument is showing that the series of positive terms
\begin{equation}\label{series1}
\sum_{j=2}^\infty \sum_{k=1}^{j-1}\frac{k(j-k)}{j^4}f_0^2\left(\frac{k}{j}\right)\,,\quad\sum_{j=2}^\infty \sum_{k=1}^{j-1}\frac{k(j-k)}{j^4}f_\pi^2\left(\frac{k}{j}\right)\,,
\end{equation}
diverge, whereas
\begin{equation}\label{series2}
\sum_{j=2}^\infty \sum_{k=1}^{j-1}\frac{k(j-k)}{j^4} f_0\left(\frac{k}{j}\right)f_\pi\left(\frac{k}{j}\right)e^{ik(\gamma_1-\gamma_2)+i\gamma_2j}
\end{equation}
converges. Under these circumstances, the only way to guarantee the convergence of \eqref{seriesimple} is to have $V_0=V_\pi=0$ as, otherwise, \eqref{series2} cannot compensate the other two, positive and divergent terms involving \eqref{series1}.

As we have shown in appendix \ref{appendix_A} we have $f_0, f_\pi\in C^\infty[0,1]$ and, hence, both functions are bounded above and below in the closed interval $[0,1]$. This immediately implies the existence of constants $M_0>0$ and $M_\pi>0$ such that
\[\sum_{j=2}^\infty \sum_{k=1}^{j-1}\frac{k(j-k)}{j^4}f_0^2\left(\frac{k}{j}\right)>M_0\sum_{j=2}^\infty \sum_{k=1}^{j-1}\frac{k(j-k)}{j^4} \,,\quad\sum_{j=2}^\infty \sum_{k=1}^{j-1}\frac{k(j-k)}{j^4}f_\pi^2\left(\frac{k}{j}\right)>M_\pi\sum_{j=2}^\infty \sum_{k=1}^{j-1}\frac{k(j-k)}{j^4}\,.\]
Now, as
\[
\sum_{j=2}^\infty\sum_{k=1}^{j-1}\frac{k(j-k)}{j^4}=\frac{1}{6}\sum_{j=2}^\infty\frac{j^3-j}{j^4}
\]
diverges we conclude that  the series \eqref{series1} diverge too.

Let us consider now the series \eqref{series2} and write it as
\[
\sum_{j=2}^\infty \frac{e^{ij\gamma_2}}{j^4}\sum_{k=1}^{j-1}k(j-k)e^{ik(\gamma_1-\gamma_2)}h\left(\frac{k}{j}\right)\,,
\]
with $h=f_0f_{\pi}$. The preceding series has the form
\[
\sum_{j=2}^\infty e^{i\gamma_2 j} A_j\,,
\]
with
\[
A_j=\frac{1}{j^4}\sum_{k=1}^{j-1}k(j-k)e^{i(\gamma_1-\gamma_2)k}h\left(\frac{k}{j}\right)\,.
\]
We use now Cauchy's convergence criterion. To this end we consider the sums
\[
\left|\sum_{j=n}^{n+p} e^{i\gamma_2 j}A_j\right|
\]
with $n\,,p\in\mathbb{N}$. Let us define
\[
S_k:=\sum_{\ell=0}^k e^{i\gamma_2\ell}\,,
\]
it is then straightforward to show that $|S_k|\leq 1/|\sin\frac{\gamma_2}{2}|$ (remember that $0<|\gamma_2|< 2\pi$). We then have
\begin{align}
   & \left|\sum_{j=n}^{n+p} e^{i\gamma_2 j}A_j\right|=\left|\sum_{j=n}^{n+p} (S_j-S_{j-1})A_j\right|=\left|\sum_{j=n}^{n+p}S_jA_j-\sum_{j=n-1}^{n+p-1}S_jA_{j+1}\right|\label{bound1} \\
   =&\left|S_{n+p}A_{n+p}-S_{n-1}A_n+\sum_{j=n}^{n+p-1}S_j(A_j-A_{j+1}) \right|\leq \frac{1}{|\sin\frac{\gamma_2}{2}|}\left(|A_{n+p}|+|A_n|+\sum_{j=n}^{n+p-1}|A_j-A_{j+1}|\right)\,.\nonumber
\end{align}
In the following we show how to find bounds for the terms appearing in the last term of \eqref{bound1}. We start with the following one for $|A_j|$
\begin{equation}\label{bound2}
|A_j|\leq\frac{1}{j^4}\sum_{k=1}^{j-1}k(j-k)\left|h\left(\frac{k}{j}\right)\right|\leq\frac{M}{j^4}\sum_{k=1}^{j-1}k(j-k)=\frac{M}{6}\frac{j^2-1}{j^3}\leq\frac{M}{6j}\,,
\end{equation}
where $M$ is an upper bound for $|h|$ in $[0,1]$. From \eqref{bound2} we immediately conclude that, as $n\rightarrow\infty$, we have
\begin{equation}\label{bound3}
|A_{n+p}|=O\left(\frac{1}{n}\right)\,, |A_n|=O\left(\frac{1}{n}\right)\,.
\end{equation}
Let us study now $\sum_{j=n}^{n+p-1}|A_{j+1}-A_j|$. In order to do this we consider
\begin{align*}
    \left|A_{j+1}-A_j\right|&=\left|\frac{1}{(j+1)^4}\sum_{k=1}^jk(j+1-k)e^{i\alpha k}h\left(\frac{k}{j+1}\right)-\frac{1}{j^4}\sum_{k=1}^{j-1}k(j-k)e^{i\alpha k}h\left(\frac{k}{j}\right)\right|\\
   &=\left|\frac{j}{(j+1)^4}e^{i\alpha j}h\left(\frac{j}{j+1}\right)+\sum_{k=1}^{j-1} \left(    \frac{k(j+1-k)}{(j+1)^4}h\left(\frac{k}{j+1}\right)-\frac{k(j-k)}{j^4}h\left(\frac{k}{j}\right) \right)e^{i\alpha k}  \right|\\
   &\leq\frac{M}{j^3}+\sum_{k=1}^{j-1}\left|\frac{k(j+1-k)}{(j+1)^4}h\left(\frac{k}{j+1}\right)-\frac{k(j-k)}{j^4}h\left(\frac{k}{j}\right)\right|\,.
\end{align*}
(here and in the following $\alpha:=\gamma_1-\gamma_2$). Now, by using the mean value theorem (remember that $h\in C^\infty([0,1])$), we have
\[
h\left(\frac{k}{j}\right)=h\left(\frac{k}{j}-\frac{k}{j+1}+\frac{k}{j+1}\right)=h\left(\frac{k}{j+1}+\frac{k}{j(j+1)}\right)=h\left(\frac{k}{j+1}\right)+h'(\sigma_{jk})\frac{k}{j(j+1)}\,,
\]
with $\sigma_{jk}\in(0,1)$, so that
\begin{align}
\left|A_{j+1}-A_j\right|\leq&\frac{M}{j^3}+\sum_{k=1}^{j-1}\left(\left|h\left(\frac{k}{j+1}\right)\left(\frac{k(j+1-k)}{(j+1)^4}-\frac{k(j-k)}{j^4}\right)\right|+|h'(\sigma_{jk})|\frac{k^2(j-k)}{j^5(j+1)}\right)\nonumber\\
\leq &\frac{M}{j^3}+M\sum_{k=1}^{j-1}\left|\frac{k(j+1-k)}{(j+1)^4}-\frac{k(j-k)}{j^4}\right|+\tilde{M}\sum_{k=1}^{j-1}\frac{k^2(j-k)}{j^5(j+1)}\leq\frac{\tilde{N}}{j^2}\,.\label{inequality}
\end{align}
where $\tilde{M}$ is an upper bound of $|h'|$ in $[0,1]$ and $\tilde{N}$ is a positive constant. We have now
\[
\sum_{j=n}^{n+p-1}|A_j-A_{j+1}|\leq\sum_{j=n}^{\infty}|A_j-A_{j+1}|=O\left(\frac{1}{n}\right)\,.
\]
Finally, plugging this and \eqref{bound3} in \eqref{bound1} we see that as $n\rightarrow\infty$
\[
\left|\sum_{j=n}^{n+p} e^{i\gamma_2 j}A_j\right|=O\left(\frac{1}{n}\right)\,,
\]
and, hence,
\[
\sum_{j=n}^\infty e^{i\gamma_2 j}A_j
\]
converges.

\end{appendices}

%
%

\end{document}